\documentclass[preprint,prb,aps,longbibliography]{revtex4-1}

\usepackage{graphicx}
\usepackage{dcolumn}
\usepackage{bm}
\usepackage{subfig}
\usepackage{amsmath}
\usepackage{amssymb}
\usepackage{multirow}
\usepackage{array}
\usepackage{tabularx}

\newcommand{\dd}{\mathrm{d}}
\begin{document}

\title{Photoacoustic generation by a gold nanosphere: From linear to nonlinear thermoelastics in the long-pulse illumination regime}

\author{Amaury Prost}
\author{Florian Poisson}
\author{Emmanuel Bossy}%
\email{emmanuel.bossy@espci.fr}

\affiliation{ESPCI ParisTech, PSL Research University, CNRS, INSERM, Universit\'e Pierre et Marie Curie, Universit\'e Paris Diderot, Institut Langevin, 1 rue Jussieu, 75005 Paris, France}

\begin{abstract}
We investigate theoretically the photoacoustic generation by a gold nanosphere in water in the thermoelastic regime.  Specifically, we consider the long-pulse illumination regime, in which the time for electron-phonon thermalisation can be neglected and photoacoustic wave generation arises solely from the thermo-elastic stress caused by the temperature increase of the nanosphere or its liquid environment. Photoacoustic signals are predicted based on the successive resolution of a thermal diffusion problem and a thermoelastic problem, taking into account the finite size of the gold nanosphere, thermoelastic and elastic properties of both water and gold, and the temperature-dependence of the thermal expansion coefficient of water. For sufficiently high illumination fluences, this temperature dependence yields a nonlinear relationship between the photoacoustic amplitude and the fluence. For nanosecond pulses in the linear regime, we show that more than $90\ \%$ of the emitted photoacoustic energy is generated in water, and the thickness of the generating layer around the particle scales close to the square root of the pulse duration. The amplitude of the photoacoustic wave in the linear regime are accurately predicted by the point-absorber model introduced by Calasso et al. [], but our results demonstrate that this model significantly overestimates the amplitude of photoacoustic waves in the nonlinear regime. We therefore provide quantitative estimates of a critical energy, defined as the absorbed energy required such that the nonlinear contribution is equal to that of the linear contribution. Our results suggest that the critical energy scales as the volume of water over which heat diffuses during the illumination pulse. Moreover, thermal nonlinearity is shown to be expected only for sufficiently high ultrasound frequency. Finally, we show that the relationship between the photoacoustic amplitude and the equilibrium temperature at sufficiently high fluence reflects the thermal diffusion at the nanoscale around the gold nanosphere.
\end{abstract}

\maketitle
\tableofcontents

\section{Introduction}
Photoacoustic imaging is a promising modality for biomedical applications that has emerged during the last two decades~\citep{beard2011,wang2012}. This non-invasive modality is based on the conversion of absorbed light energy into ultrasound via the thermoelastic effect~\citep{gusev1993}. The image contrast therefore depends on the optical absorption properties of the medium. In biological tissues, endogenous optical absorption may be used to form various types of images. For instance imaging of the hemoglobin enables reconstruction of the vascularization network \citep{zhang2009}. To further enhance the contrast or obtain complementary information, various exogenous contrast agents have been developed for photoacoustics~\citep{wu2014}. 
In particular, plasmonic noble metal nanoparticles have been introduced as photoacoustic contrast agents in the early 2000s~\citep{oraevsky2001,eghtedari2003}.  Gold nanoparticles (GNP) are very attractive as photoacoustic contrast agents thanks to their large optical absorption cross-section, their resistance to high illumination fluences and their spectral selectivity based on their surface plasmon resonance~\citep{jain2006}. 
Typically, the optical absorption cross-section of noble metal nanoparticles is a few orders of magnitude larger than that of traditional molecular dyes~\citep{jain2006,wu2014}. One major consequence of the strong optical cross-section of GNPs is that their temperature can increase significantly when they absorbe pulsed light~\citep{pustovalov2005}, leading to various possible phenomena including nano/micro-bubble formation~\citep{egerev2008,pustovalov2008,Lapotko2009,zharov2011} or even modifications of the particle shape. In the context of photoacoustic imaging, bubble formation is interesting as the emitted signals are usually strong and exhibit a nonlinear relationship between the incident fluence and the photoacoustic amplitude~\citep{Sarimollaoglu2014,zharov2011}. More generally, nonlinear photoacoustic phenomena provide a means of selectively detecting contrast agents from an absorption background that behaves linearly~\citep{Sarimollaoglu2014}, similarly to what is done in the field of ultrasound imaging. Several phenomena may induce nonlinear relationships between the photoacoustic signal amplitude and the energy of the incident light in addition to bubble formation, such as optical saturation~\citep{danielli2010,zharov2011}, photochemical reaction~\citep{OConnor1983} and temperature-dependent thermodynamic parameters~\citep{diebold2001,burmistrova1979,gusev1993,inkov2001,egerev2008,Danielli2014,Wang2014,simandoux2014}. The latter phenomenon is related to the temperature dependence of the thermal expansion coefficient, and is at the core of this paper, in which we investigate theoretically the photoacoustic generation by a gold nanosphere in the thermoelastic regime (absence of bubble formation). In this work, our study is restricted to what we call the long-pulse illumination regime (typically longer than several picoseconds), for which the typical electron-phonon relaxation time in the gold nanoparticle is negligible compared to the illumination duration. In particular, electrons and phonon may be considered not only to have thermalized distributions (such as assumed in the two-temperature model~\citep{eesley1983}), but also to be described by a unique temperature value, a situation very different from that encountered with sub-picosecond light pulses where nonthermal distributions may be involved~\citep{tas1994,groeneveld1995}. In the long-pulse illumination regime, a single value for temperature may be used to describe thermodynamic properties, and the photoacoustic wave generation arises from the thermo-elastic stress caused by the temperature increase of the nanosphere or its liquid environment. This regime encompasses the nanosecond-pulse regime commonly used in photoacoustic imaging, and to which most of our study of thermal nonlinearity is restricted.

Although nanoparticles have now been used for more than a decade as contrast agents for photoacoustic imaging in many research studies, comparatively few studies investigated the physics involved at the scale of the nanoparticle, whether with theoretical or experimental approaches. 

In the linear regime, ~\citet{inkov2001} introduced a three-step model to predict the photoacoustic emission by an absorbing spherical particle, based on solving (1) a light absorption problem, (2) a thermal diffusion problem and (3) an acoustic problem. Analytical solutions were provided in the linear regime, but were limited to thermally small or large particles, assumptions that are not valid for gold nanospheres illuminated with nanosecond pulses, as commonly encountered in photoacoustic imaging. For nanospheres illuminated with nanosecond pulses, the thermal relaxation time is comparable to the pulse duration, making it complex to analytically derive the temperature field. Numerical approaches are usually required to solve the thermal problem in this case, as was done for instance by~\citet{baffou2011}. More recently, it has been demonstrated experimentally in  the linear regime that for gold nanospheres illuminated with nanosecond pulses, it is mostly the liquid surrounding the nanoparticles that emits photoacoustic waves~\citep{chen2012}, as was theoretically discussed earlier for thermally small particles~\citep{inkov2001}. A theoretical explanation in the linear regime was proposed by~\citet{chen2012}, but the temperature field was modeled via a quasi-static thermal field. Other studies provided theoretical expressions for the photoacoustic emission by spherical absorbers, but with assumptions not always valid in this work and more importantly also limited to the linear regime~\cite{diebold1988,diebold2002}. Two recent studies also reported comparisons between experimental results and theoretical predictions in the linear regime~\citep{shinto2013,fukasawa2014}, but the authors assumed heat and stress confinement at the scale of a whole particle suspension and did therefore not consider the photoacoustic generation at the scale of individual particles.

In the nonlinear thermoelastic regime, a few works have been reported with nanoparticles. ~\citet{diebold2001} provided analytical expressions of the waveforms emitted by a point-absorber model. In our work, we extensively use the predictions of this model to be compared to our own model that takes into account the finite size of the gold nanosphere, both in the linear and nonlinear regime. Following the physical approach introduced by~\citet{inkov2001}, ~\citet{egerev2008} described and observed the nonlinear photoacoustic generation by gold nanospheres in water in the thermoelastic regime. A simple criteria to assess the significance of nonlinear generation was derived based on scaling arguments, that shows that thermal nonlinearity should be observed at sufficiently high fluences (Eq. (3) of ~\citet{egerev2008}), as demonstrated experimentally. The validity of this criteria will be discussed in sections~\ref{subsub:PredictionFromNonLinearAnalyticPointModel} and \ref{subsubsec:optimal size for nonlinear generation}, in comparison with the quantitative predictions from both the point-absorber model and our model. In the thermoelastic regime, another previous experimental study reported a nonlinear increase of the photoacoustic signal with the laser fluence~\citep{Nam2012}. The origin of the nonlinearity was assumed to be the temperature-dependence of the thermal expansion coefficient, but presumably caused by thermal coupling within aggregated nanoparticles in cells, rather than by temperature elevation around individual nanoparticles as considered in our work.

The main objective of our theoretical work is to provide physical insight and quantitative predictions regarding the photoacoustic generation by a single gold nanosphere, for both the linear and nonlinear thermoelastic regimes, beyond the few initial results from previous relevant studies~\citep{diebold2001,inkov2001,egerev2008,chen2012}. In section~\ref{sec:TheoreticalApproach}, we first introduce the physical model used to predict the photoacoustic signal from a gold nanosphere. Our model takes into account the finite size of the nanosphere and its elastic and thermo-elastic properties, as well as the temperature-dependence of the thermal expansion coefficient of the surrounding liquid. Predictions from the point-absorber model are also given for further comparison in the results section. Section~\ref{sec:TheoreticalApproach} also describes the principles that were used for solving the thermal and acoustic problems, with details further given in Appendixes~\ref{appendix:GreenTh} and ~\ref{appendix:FDTD}. Section~\ref{sec:Results} provides results and discussions in both the linear and nonlinear thermoelastic regimes. In the linear regime, we study the origin of the photoacoustic wave as a function of the size of the gold nanosphere and the pulse duration. From there, the paper focuses on the nanosecond pulse regime for which the generation mostly occurs in the surrounding liquid. A scaling law is found that describes the typical thickness of the water layer that generates the photoacoustic wave. A comparison between our results and those from the point-absorber model~\citep{diebold2001} is then provided, as a preamble to further analysis in the nonlinear regime. In the nonlinear regime, we first derive predictions from the expressions given by the point-absorber model. We then describe quantitatively temperature rises in gold nanospheres, showing that thermal nonlinearity is indeed expected at commonly encountered light fluences. Our predictions for the photoacoustic amplitude emitted by a gold nanosphere in the nonlinear regime are given and compared to those from the point-absorber model, and we study the occurence of nonlinearity as a function of fluence and particle size. Finally, we investigate the influence of the equilibriume temperature on the photoacoustic amplitude, for a given nanosphere and fixed fluence.

\section{Theoretical approach}\label{sec:TheoreticalApproach}
We consider the photoacoustic generation by a single gold nanosphere in water. In this section, we describe the models and computations that are used to produce the theoretical predictions given in Sec.~\ref{sec:Results}. Throughout all this work, spherical symmetry is assumed, with the center of the gold nanosphere as the center of symmetry.

\subsection{Physical models}
\label{sub:PhysicalModel}

\subsubsection{Photoacoustic generation by a point absorber in a liquid}
\label{subsub:PAFluidModel}

In order to discuss further below the generation of photoacoustic waves by a gold nanosphere and to introduce relevant quantities used throughout our work,  we first recall here the basic equations which describe the photoacoustic generation in the simple situation of a mechanically and thermally homogeneous liquid medium. In particular, we provide the analytical expressions derived by \citet{diebold2001} of the photoacoustic pressure wave emitted for the limiting case of a point-absorber, both in the linear and nonlinear thermo-elastic regime. The corresponding analytical predictions  are discussed in the results section, in particular in comparison with the theoretical predictions for a gold nanosphere. We will show that the point-absorber model accurately predicts the photoacoustic emission by a gold nanosphere only in the linear regime, which justifies the introduction of our nonlinear model for the gold nanosphere in the next section.

\paragraph{Model equations.} When the physical properties are assumed to be homogeneous and  constant in time, the generation of photoacoustic waves in  a liquid medium is dictated by the following system of coupled equations~\citep{morse1986,gusev1993,diebold2001}
\begin{align}
\rho_0 c_p\frac{\partial T}{\partial t}(\mathbf{r},t)-\kappa\Delta T (\mathbf{r},t)=P_v(\mathbf{r},t) \label{eq:ThermalEqHomogeneous}\\
\Delta p(\mathbf{r},t) -\frac{1}{c_s^2}\frac{\partial^2 p}{\partial t^2}(\mathbf{r},t)=-\rho_0\beta_0\frac{\partial^2 T}{\partial t^2} (\mathbf{r},t)
\label{eq:PAEqHomogeneous}
\end{align}
where $p(\mathbf{r},t)$ and $T(\mathbf{r},t)$ are respectively the (photo)acoustic pressure field and the temperature field. The relevant physical properties are the mass density $\rho_0$, the coefficient of thermal expansion  $\beta_0=-\frac{1}{\rho_0}\left(\frac{\partial \rho}{\partial T}\right)_0$, the acoustic velocity $c_s$, the thermal conductivity $\kappa$ and the specific heat capacity at constant pressure $c_p$. Eq.(\ref{eq:ThermalEqHomogeneous}) is a standard heat diffusion equation, with a heat source term $P_v(\mathbf{r},t)$ representing the volumetric density of power converted to heat (with dimensions unit power per unit volume). Eq.(\ref{eq:ThermalEqHomogeneous}), as fully decoupled from the pressure field, is only valid for liquid and solid media (as opposed to gases) for which the ratio $\gamma = \frac{c_p}{c_v}$ of the specific heat capacity at constant pressure to the specific heat capacity at constant volume  may be considered very close to 1 (see demonstration in Appendix~\ref{appendix:DerivationEquation} based on~\citet{morse1986}), an assumption that will be made throughout all this work. The photoacoustic wave equation (\ref{eq:PAEqHomogeneous}) is a classical wave equation with a source term given by the second time-derivative of the temperature field.  

Equations (\ref{eq:ThermalEqHomogeneous}) and (\ref{eq:PAEqHomogeneous}) indicate that to solve the photoacoustic problem given a source term $P_v(\mathbf{r},t)$, one has to \textit{first} solve a thermal diffusion problem, and \textit{then} solve an acoustic problem once the source term given by the temperature field is known. In the context of photoacoustics, the heat source term arises from optical absorption, and is therefore proportional to some illumination function such as the fluence rate (or intensity) $\Phi_r(\mathbf{r},t)$ (unit power per unit surface). In this work, we will considered a single optical absorber (with an absorption cross-section $\sigma_a$) illuminated with some incident pulsed light described by the following expression
\begin{equation}
\Phi_r(\mathbf{r},t)=\Phi_0 \frac{1}{\tau_p} f(\frac{t}{\tau_p})
\end{equation}
where $\Phi_0$ is the fluence (unit energy per unit surface) and $f$ is a dimensionless peaked function describing the temporal profile of the fluence rate. $f$ verifies $\int_{-\infty}^{+\infty} f(\hat{\tau})\mathrm{d}\hat{\tau}=1$ and is normalized such as $\tau_p$ is defined as the full width at half maximum ($\tau_p$ is further referred to in the text as the pulse duration). Throughout this work, the laser temporal profile is chosen as a Gaussian defined accordingly by 
\begin{equation} \label{eq:GaussianPulse}
f(\hat{\tau})=\frac{2 \sqrt{\ln(2)}}{\sqrt{\pi}} e^{-4 \ln(2)\hat{\tau}^2}
\end{equation}

\paragraph{Point absorber in the linear regime.} For an optical absorber of vanishingly small size, but with a finite optical absorption cross-section $\sigma_a$, \citet{diebold2001} provided an analytical expression of the photoacoustic pressure wave emitted by the "point-absorber" (referred to as a photoacoustic point source in \citet{diebold2001}):
\begin{equation}
\label{eq:AnalyticDieboldLinear}
p(\mathbf{r},t)=E_{abs}\beta_0  \frac{1}{c_p \tau_p^2}  \frac{1}{4 \pi r}\frac{\dd f}{\dd \hat{\tau} }\left(\hat{\tau}=\frac{t-\frac{r}{c_s}}{\tau_p}\right)
\end{equation}
where $E_{abs}=\sigma_a \Phi_0 $ is the energy absorbed by the point absorber. As expected from the linearity of Eqs. (\ref{eq:ThermalEqHomogeneous}) and (\ref{eq:PAEqHomogeneous}), the photoacoustic pressure is proportional to the absorbed optical energy. 

\begin{figure}[h!]
\includegraphics{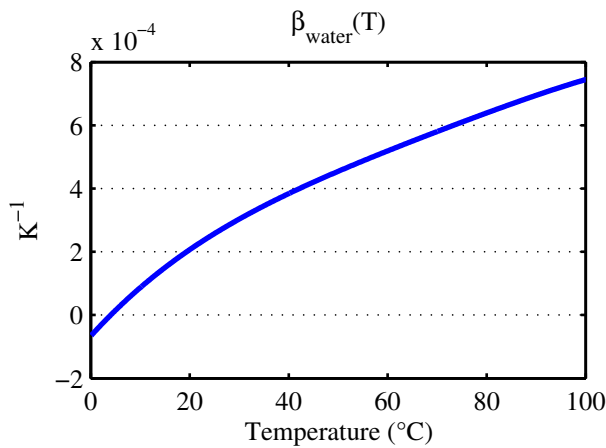} 
\caption{\label{fig:Beta} Thermal expansion coefficient of water as a function of temperature. Data derived from the density of water as a function of temperature~\citep{handbook}}.
\end{figure}
\paragraph{Nonlinear thermo-elastic regime.} When significant temperature rises occur, the physical properties involved in (\ref{eq:ThermalEqHomogeneous}) and (\ref{eq:PAEqHomogeneous}) may vary during the illumination and subsequent photoacoustic generation. It is well known that amongst the relevant thermodynamics properties, the thermal expansion coefficient $\beta$ shows the most significant temperature dependency~\citep{gusev1993}. The temperature dependence of the thermal expansion coefficient $\beta(T)$ of water is shown in Fig.~\ref{fig:Beta}. Taking into account this temperature dependence, Eq.~\ref{eq:PAEqHomogeneous} has to be modified as  
\begin{equation}\label{eq:NonLinearPAfluid}
c_s^2 \Delta p(\mathbf{r},t)-\frac{\partial^2 p}{ \partial t^2}(\mathbf{r},t)=-\rho_0 c_s^2\frac{\partial}{\partial t}\left(\beta(T)\frac{\partial T}{\partial t}(\mathbf{r},t)\right)
\end{equation}
Whereas the temperature field remains linearly related to the optical illumination (via Eq.~\ref{eq:ThermalEqHomogeneous}), the photoacoustic pressure wave in Eq.~\ref{eq:NonLinearPAfluid} is nonlinearly dependent on the temperature field and therefore on the optical illumination. Note that the source term in Eq.~\ref{eq:NonLinearPAfluid} is slightly different from that given initially in the pioneer work by \citet{burmistrova1979} or in~\citet{gusev1993} (see detailed derivation of our equation in Appendix~\ref{appendix:DerivationEquation}).

\paragraph{Point absorber in the nonlinear regime.} For the point absorber model, under the assumption that the temperature dependence of the thermal expansion coefficient can be linearized as $\beta(T)=\beta_0+\beta_1 (T-T_0)$ (with $\beta_0=\beta(T_0)$ and $\beta_1=\frac{\dd \beta}{\dd T}(T_0)$), and for a gaussian temporal profile of the illumination pulse, \citet{diebold2001} also provided an analytic expression of the photoacoustic pressure wave emitted by the point absorber in the nonlinear regime:
\begin{align}\label{eq:CalassoNonlinearPA}
p(\mathbf{r},t)=&E_{abs}\beta_0  \frac{1}{c_p \tau_p^2} \frac{1}{4 \pi r} \frac{\dd f}{\dd \hat{\tau} }\left(\hat{\tau}=\frac{t-\frac{r}{c_s}}{\tau_p}\right) + \nonumber \\
& E_{abs}^2\beta_1\frac{1}{\rho_0 \chi^{3/2} c_p^2 \tau_p^{7/2}}\frac{1}{4 \pi r}h\left(\hat{\tau}=\frac{t-\frac{r}{c_s}}{\tau_p}\right)
\end{align}
where $h(\hat{\tau})$ is a dimensionless function with a tripolar shape, given in details by Eq.~(\ref{eq:h_function}) in Appendix~\ref{appendix:DerivationCalasso}. Note that the numerical prefactor in the nonlinear term of Eq.~\ref{eq:CalassoNonlinearPA} is different from that the original equation (25) given in \citet{diebold2001}, for various reasons detailed in Appendix~\ref{appendix:DerivationEquation}, including the modification required to take into account the correct source term of Eq.~\ref{eq:NonLinearPAfluid}.

\subsubsection{Photoacoustic generation by a gold nanosphere in a liquid}
\label{subsub:PAGNSModel}

In this section, we present the physical models used to describe the photoacoustic emission by a finite-size gold nanosphere (of radius $R_s$) immersed in water. As opposed to the case of a point absorber, no analytical expression is available for the emitted photoacoustic pressure wave by a finite-size solid sphere, except for very limiting cases in the linear regime~\citep{diebold1988, diebold2002, diebold2002, egerev2009}, out of scope here. In \citet{chen2012}, the thermal source term was modelled via a quasi-static thermal field, and the solution in the linear regime was approximated in the Fourier domain assuming the sphere was small compared to the ultrasound wavelength. The model used in our work 
takes into account thermal diffusion around the nanosphere, photoacoustic generation and propagation in both the gold nanosphere and its liquid environment, and any arbitrary temperature-dependence of the thermal expansion coefficient of the liquid environment.

\paragraph{Thermal model.} Because the thermal conductivity of gold is much larger than that of water, the temperature within the nanosphere is considered in this work to be uniform, which is known to be a very accurate approximation  for gold spheres of diameter of the order of a few tens of nanometers, and for pulse duration no shorter than a few ps~\citep{baffou2011,pustovalov2005}. In particular, we emphasize that for the pulse durations considered in our work, longer than a few tens of ps, the electron-phonon thermalization that occurs on a time scale no longer than a few ps can be totally neglected~\citep{baffou2011,pustovalov2005}, as opposed to the situation usually encountered in picosecond acoustics with sub-picosecond illumination~\citep{tas1994,groeneveld1995}. Under this assumption, the spatio-temporal evolution of the temperature fields $T_s(t)$ inside the solid gold nanosphere and $T_f(\mathbf{r},t)$ in its liquid environment can be described by the following system of differential equations~\cite{baffou2011}:

\begin{subequations}\label{eq:ThermalEqs}
\begin{align}
\frac{\partial T_{s}}{\partial t}(t)-
\frac{3}{R_s}\frac{\kappa^f}{\rho_0^{s} c_p^{s}}\frac{\partial T_f}{\partial r}(R_s^+,t)
= &\ \frac{\sigma_{abs}\Phi_0}{\rho_0^{s} c_p^{s}\frac{4}{3}\pi R_s^3}\frac{1}{\tau_p} f(\frac{t}{\tau_p}) 
 \label{eq:TsEquation}\\
\frac{\partial T_f}{\partial t}(\mathbf{r},t)-\frac{\kappa^f }{\rho_0^f c_p^f }\Delta T_f(\mathbf{r},t)=&\ 0, \; r>R_s 
\end{align}
\end{subequations}
with the following boundary conditions:
\begin{subequations}\label{eq:TBoundaryCond}
\begin{align}
T_f(R_s^+,t)&=  T_s(t) \label{eq:ContinuityT}\\
T_f(r \to \infty ,t)&=T_{0} \label{eq:Tinfty}
\end{align}
\end{subequations}
with $r=\|\mathbf{r}\|$, and the subscripts $s$ and $f$ referring respectively to the solid and fluid phases. Eq. \ref{eq:TsEquation} states that the variation of the uniform sphere temperature increases via the absorbed optical energy and decreases via thermal conduction at the gold/water interface. This equation takes into account the continuity condition for the thermal flux across the interface. Eqs. \ref{eq:ContinuityT} and \ref{eq:Tinfty} provide the additional boundary conditions required to solve the problem. The continuity equation Eq. \ref{eq:ContinuityT} assumes that any interfacial thermal resistivity is neglected. This assumption is discussed further in section~\ref{subsec:Discussion}.

\paragraph{Thermoelastic equations. } 

Under spherical symmetry and for isotropic materials, the thermoelastic equations in both heterogeneous solid and liquid media can be written as a first order velocity-stress system of equations that reads~\citep{royer1999,chen2012}

\begin{subequations}\label{eq:SolidPAEq}
\begin{align}
\frac{\partial v_r}{\partial t}(\mathbf{r},t)=& +\frac{1}{\rho_0(r)}\left[\frac{\partial \sigma_{rr}}{\partial r}(\mathbf{r},t) + \frac{2}{r}(\sigma_{rr}-\sigma_{\theta \theta})\right] \\
\frac{\partial \sigma_{rr}}{\partial t}(\mathbf{r},t)=& \left[ (\lambda(r)+2\mu(r)) \frac{\partial }{\partial r}+2 \lambda(r)\frac{1}{r}\right]v_r(\mathbf{r},t) \nonumber\\
&-(\lambda(r)+\frac{2}{3}\mu(r))\beta(T(\mathbf{r},t)) \frac{\partial T}{\partial t}(\mathbf{r},t) \\
\frac{\partial \sigma_{\theta \theta}}{\partial t}(\mathbf{r},t)=& \left[ \lambda(r)\frac{\partial }{\partial r}+2 (\lambda(r)+\mu(r))\frac{1}{r}\right]v_r(\mathbf{r},t) \nonumber\\
&-(\lambda(r)+\frac{2}{3}\mu(r))\beta(T(\mathbf{r},t)) \frac{\partial T}{\partial t}(\mathbf{r},t)
\end{align}
\end{subequations}
where $\mathbf{\sigma}$ is the stress tensor, $v_r$ is the radial displacement velocity, and $\lambda$ and $\mu$ are the Lam\'{e} coefficients. One can readily verify that if $\mu$ is set to zero in Eqs. \ref{eq:SolidPAEq}, i.e. the material is a liquid ($\sigma_{rr}=-p$), the system yields Eq. \ref{eq:NonLinearPAfluid} (or Eq. \ref{eq:PAEqHomogeneous} for constant $\beta$). In the relevant case here of a solid/liquid interface, the following continuity conditions must hold for both the radial velocity  and stress at the sphere interface :
\begin{subequations}\label{eq:ContinuitySolidPAEq}
\begin{align}
v_r(R_s^-,t)&=v_r(R_s^+,t)\\
\sigma_{rr}(R_s^-,t)&=-p(R_s^+,t)
\end{align}
\end{subequations}

\subsection{Computations for a gold nanosphere}\label{sub:Computations}
The equations that describe the photoacoustic generation by a solid and optically absorbing sphere (Eqs. \ref{eq:ThermalEqs} to \ref{eq:ContinuitySolidPAEq}) are much more complex than the equations for a homogeneous liquid (Eqs. \ref{eq:ThermalEqHomogeneous} to \ref{eq:PAEqHomogeneous}) and cannot be solved analytically. However, their resolution still requires to first compute the temperature field from the thermal problem, and then to use this temperature field as a source term in the thermoelastic problem. The full resolution of both the thermal and thermoelastic problems is referred to further in the text as a numerical simulation, based on the computational approaches described below.

\subsubsection{Temperature computations}\label{subsub:ThermalSolution}
The system of equations \ref{eq:ThermalEqs} and \ref{eq:TBoundaryCond} may be solved analytically for an impulse excitation, i.e $\frac{1}{\tau_p} f(\frac{t}{\tau_p})\to\delta(t) $, using the Laplace Transform. After tedious but simple algebric manipulations and use of tables of known inverse Laplace transforms, one may obtain the Green's function $G_{\mathrm{th}}(\mathbf{r},t)$ (solution to a $\delta(t)$ excitation) of the thermal problem, as was done by~\citet{egerev2009}. The expression of $G_{\mathrm{th}}(\mathbf{r},t)$ is given in Appendix~\ref{appendix:GreenTh}. The temperature field in water for a pulse excitation can then be calculated by the convolution of the thermal Green's function with the source term:
\begin{equation}
\label{eq:SolutionT}
T(\mathbf{r},t)= G_{\mathrm{th}}(\mathbf{r},t)\ast \frac{\sigma_{abs}\Phi_0}{\rho_0^{s} c_p^{s}\frac{4}{3}\pi R_s^3}\frac{1}{\tau_p}f(\frac{t}{\tau_p})
\end{equation}
For all our results, the convolution in Eq.~\ref{eq:SolutionT} was performed numerically, with the source function $f$ given by Eq.~\ref{eq:GaussianPulse}. The temperature field $T(\mathbf{r},t)$ was computed and sampled on a regular grid $T(n\times\Delta r,m\times \Delta t)$ required by the finite-difference in time-domain resolution of the thermoelastic problem described below.

\subsubsection{Acoustic computations}\label{subsub:ElasticSolution}

In this work, we used a finite-difference time-domain (FDTD) algorithm to solve the thermo-elastic problem. We adapted the well-known Virieux's scheme to our problem with spherical symmetry. In brief, the Virieux's scheme for elastodynamics~\citep{virieux1986} (analog to the Yee's scheme for electromagnetism~\citep{yee1966}) is based on a spatio-temporal discretization of the system of continuous equations (Eqs. \ref{eq:SolidPAEq}) on staggered grids (spatial grid step $\Delta r$ and temporal grid step $\Delta t$). The solution is computed step by step in time, over the whole spatial domain at each time step. Any known source term may be taken into account, both in the sphere and in water. In particular, it makes it straightforward to take into account the temperature-dependence of the thermal expansion coefficient of water, by simply computing the value of $\beta^f(T_f(\mathbf{r},t))$ at each point in space and time. Another well-known key advantage of the Virieux's scheme is that boundary conditions such as given by Eqs.~\ref{eq:ContinuitySolidPAEq} are implicitly verified~\cite{virieux1986}. As a consequence, reflected and transmitted acoustic waves  at the water-gold interface were taken into account in our numerical solutions. The discretized equations that were used are detailed in Appendix~\ref{appendix:FDTD}. The spatial grid step $\Delta r$ was chosen small enough to ensure a proper convergence of the FDTD solution: the convergence was ensured by verifying that results with two different spatial steps showed no significant difference. The values of $\Delta r$ typically ranged from 0.1 nm to 5 nm depending on the sphere radius and the pulse duration. The dimension of the spatial domain was taken sufficiently large (typically several tens of $\mu$m) such that any spurious reflections from the domain boundary  would arrive far after the photoacoustic pressure waveforms. The time step $\Delta t$ was derived from $\Delta r$ via the stability condition given in Appendix~\ref{appendix:FDTD}.

\subsubsection{Values of the physical properties}\label{subsub:PhysicalProperties}

\begin{table}[]
		\caption{Physical constants associated with gold and water, at T $\sim 25 ^\circ C$, from \cite{handbook}}
\begin{tabular}{lccc}
\hline
\hline
Properties & Gold & Water & Unit\\
\hline
Mass density $\rho_0$ & 19.3 &1.00&$\times 10^3\mathrm{kg.m}^{-3}$ \\
Specific heat capacity $c_p$  & 129 &4200&$\mathrm{J.kg}^{-1}\mathrm{.K}^{-1}$ \\
Thermal conductivity $\kappa$ & 318 &0.60&$\mathrm{W.m}^{-1}\mathrm{.K}^{-1}$ \\
Thermal diffusivity $\chi$ & 128 &0.142&$\times 10^{-6}\mathrm{m}^{2}\mathrm{.s}^{-1}$ \\
Thermal expansion $\beta$ & 0.43 & Fig.\ref{fig:Beta}&$\times 10^{-4}\mathrm{K}^{-1}$ \\
First Lam$\acute{e}$ coefficient $\lambda$ & 147 & 2.25 & GPa\\
Second Lam$\acute{e}$ coefficient $\mu$ & 27.8 & 2.25 & GPa\\
Compressional wave velocity & 3.24 &1.50& $\mu\mathrm{m.}\mathrm{ns}^{-1}$\\
Shear wave velocity & 1.20 &-& $\mu\mathrm{m.}\mathrm{ns}^{-1}$\\
\hline
\hline
\end{tabular}
\label{tab:MaterialProperties}
\end{table}

All the values of the physical properties of gold and water used in the computations are summarized in Table  \ref{tab:MaterialProperties}. Except for the thermal expansion coefficient whose value may depend on temperature, the values for all other properties (assumed to be constant) were those at room temperature ($\sim 25 ^\circ C$). 
The absorption cross-section of a gold nanosphere depends on its size, and therefore so does the absorbed energy for a given fluence. The values of absorption cross-sections used in this work were derived from the Mie theory with optical constants from \citet{Johnson1972}, for an illumination wavelength $\lambda\ = \ 532 \ \mathrm{nm}$. A few typical values are given in Table~\ref{tab:Cross-Sections}. For a diameter below typically 50 nm, the absorption cross-section is much larger than the scattering cross-section and scales as the nanoparticle volume~\citep{jain2006}.

\begin{table}[h!]
\caption{Absorption cross-section $\sigma_{abs}$ of a gold nanosphere as a function of radius, for an illumination wavelength $\lambda\ = \ 532 \ \mathrm{nm}$ }
\begin{tabular}{m{90pt} c c c c c c c}
\hline
\hline
$R_s$ ($\mathrm{nm}$) &10 & 20 & 30 & 40 & 50 & 60\\
\hline
$\sigma_{abs}$ ($\times 10^{-14}~\mathrm{m^2}$) & $3.7 \times 10^{-2}$ & $3.3 \times 10^{-1}$  & $1.1$ & $1.8$ & $2.1$ & $2.3$ \\   
\hline
\hline
\end{tabular}
\label{tab:Cross-Sections}
\end{table}

\section{Results and discussion}\label{sec:Results}

\subsection{Linear regime}\label{sub:LinearRegime}
In this section, the temperature-dependence of the thermal expansion coefficient of water is neglected, i.e we consider the linear photoacoustic regime. Our first objective is to investigate the relative contribution of the gold nanosphere and its liquid environment to the sound generation. When the photoacoustic wave is predominantly generated from the liquid environment rather than from the solid sphere, we then investigate the typical thickness of the water layer that generates the photoacoustic wave. As a preamble to our results in the nonlinear regime, our results for the gold nanosphere are compared to those predicted by the point absorber model in the linear regime first. Throughout all the paper, all absolute photoacoustic amplitudes are given at 1 mm from the center of the absorber ($r=1$ mm). In the linear regime, all the results are predicted for an equilibrium temperature $T_0=20^\circ C$.

\subsubsection{Origin of the photoacoustic wave}\label{subsub:OriginPA}

\begin{figure*}[!h]
\includegraphics[width = 18 cm]{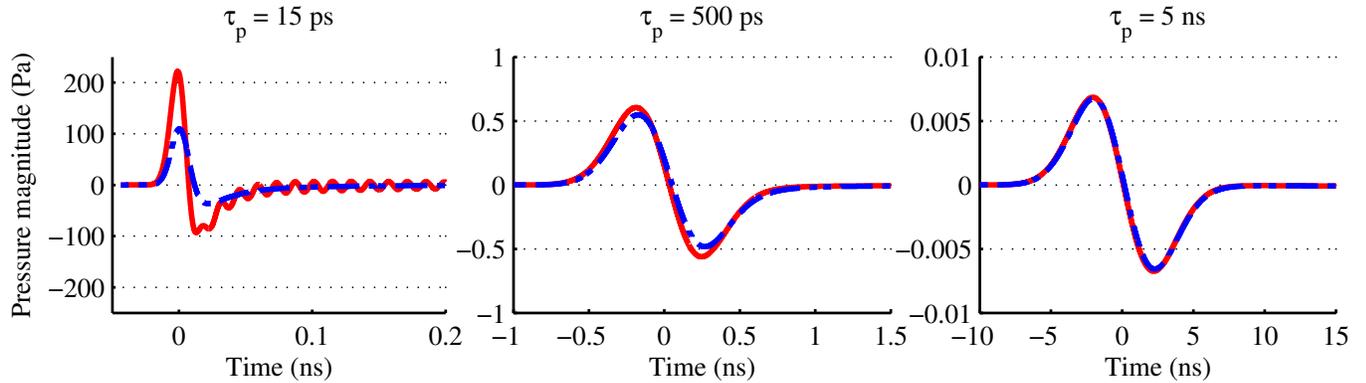} 
\caption{Photoacoustic waveforms obtained in the linear regime, for a 20-nm radius sphere, illuminated with a fluence $\Phi=1~\mathrm{mJ.cm^{-2}}$. The red curves correspond to signals obtained with thermal expansion in both water and gold (S1). The blue curves correspond to signals obtained with no thermal expansion in gold (S2).}
\label{fig:PAWavesFromSphereVsEnvironment}
\end{figure*}

The absorption of the laser pulse by a spherical gold nanoparticle creates a transient temperature rise  in both the particle and its liquid environment due to heat diffusion. From Eqs.~\ref{eq:SolidPAEq}, it is clear that both the gold nanosphere and  its environment may generate photoacoustic waves. Here, we investigate the relative contribution to the photoacoustic signal from the gold nanosphere and from its water environment, as a function of the pulse duration $\tau_p$ and sphere radius $R_s$. The considered radii are on the order of a few nanometers to tens of nanometers, and the pulse durations typically range from tens of picoseconds to tens of nanoseconds. For each pair of parameters ($\tau_p, R_s$), two different numerical simulations were performed. Simulation (S1) computed the photoacoustic wave generated from the whole system, i.e the gold nanosphere and its water environment; simulation (S2) was identical except that the thermal expansion coefficient of gold was set to zero. The results from (S2) therefore only takes into account sound generation from water. Fig.\ref{fig:PAWavesFromSphereVsEnvironment} shows a plot of the waveforms from simulations (S1) and (S2) for a 20-nm radius gold nanosphere for three different values of $\tau_p$ (10 ps, 500 ps and 5 ns). It is clear from Fig.\ref{fig:PAWavesFromSphereVsEnvironment} that the predominant origin of the photoacoustic generation highly depends on the pulse duration: "short" pulses mostly excite acoustic waves in the nanosphere, which are then radiated into the water,  whereas photoacoustic waves with "long" pulses originate mostly from the liquid around the nanosphere. Both regimes have been studied experimentally. Indeed, various investigations have been conducted on acoustic vibration of gold nanoparticles in the short pulse regime (fs or ps excitation), see for instance \citep{delfatti1999}. In the nanosecond regime, Chen et al. have experimentally demonstrated that the photoacoustic signals originate from the environment rather than the nanosphere itself~\citep{chen2012}. Their demonstration was based on the fact that the photoacoustic signal amplitude followed the properties of the temperature-dependence of the thermal expansion coefficient of the liquid around the particle. Fig.\ref{fig:PAWavesFromSphereVsEnvironment} illustrates that our model and simulations encompass these different regimes, from the excitation of vibration modes in the sphere by short pulses (although no shorter than a few picosecond as a requirement of our thermal model) to photoacoustic generation directly in the surrounding liquid. It is therefore adapted to model a variety of different phenomena.  
\begin{figure}[h!]
\includegraphics[scale=1]{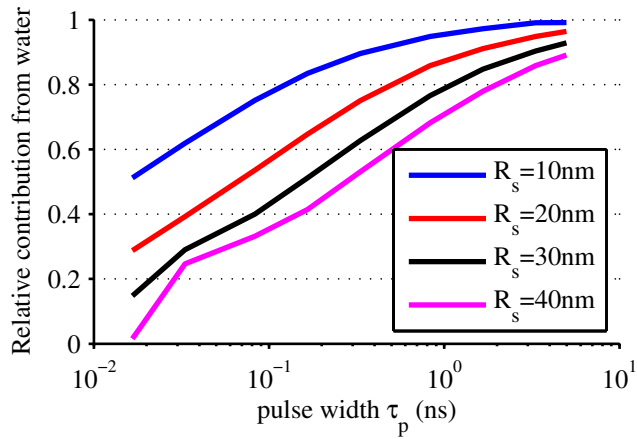}
\caption{\label{fig:RelativeContribution} Relative contribution from water to the overall photoacoustic energy (from water and gold), as a function of the laser pulse duration $\tau_p$ and the nanosphere radius $R_s$. For a pulse duration of $\tau_p=5\ \mathrm{ns}$, more than typically $90\ \%$ of the energy is emitted from water.}
\end{figure}
In addition, our model can provide a quantitative assessment of the relative contribution to the generated photoacoustic wave as a function of pulse duration $\tau_p$: to do so, the energy $\epsilon$ of the emitted photoacoustic wave (defined as $\epsilon = \frac{4 \pi r^2}{\rho_0 c_p}\int_{-\infty}^{+\infty} p^2(r,t) \dd t$, $r> R_s$, independent of $r$) was computed for simulations (S1) and (S2). The relative contribution from water was defined as $\eta=\frac{\epsilon(S2)}{\epsilon(S1)}$. The values of $\eta$ as a function of pulse duration and sphere diameter, plotted on Fig.\ref{fig:RelativeContribution}, show that both the sphere diameter and the pulse duration affect the relative contribution from water. However, for pulse durations larger than a few nanoseconds, most of the photoacoustic energy comes from the surrounding water, in agreement with the experimental results in \citep{chen2012}. For sphere diameters up to 40 nm, more than $90 \%$ of the photoacoustic energy is generated in water.\\

From this point and throughout the rest of paper, we focus our interest on the nanosecond regime, for which the photoacoustic emission from the nanosphere is negligible compared to that of water around it. Within this context, the following two sections investigate and quantify the typical dimension of the water layer that generates the photoacoustic wave, and compare the photoacoustic amplitude predicted for the nanospheres to those predicted from the point-absorber model~\citep{diebold2001}.

\subsubsection{Typical thickness of the generating water layer}\label{subsub:ExtentPALayer}

\begin{figure}[!h]
\includegraphics[scale=1]{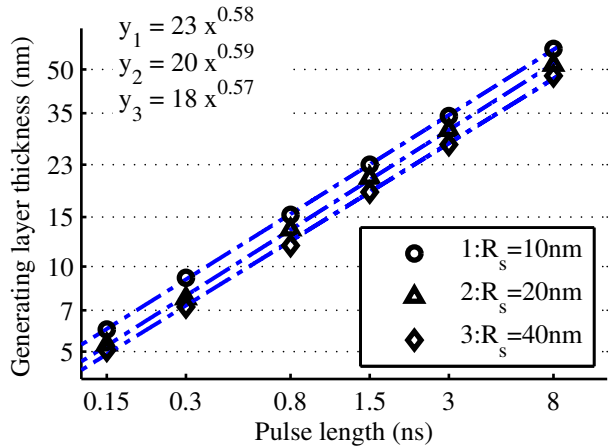} 
\caption{\label{fig:Shell_dependence} Typical thickness of the water layer that emits the photoacoustic wave as a function of the pulse duration $\tau_p$, for different values $R_s$ of the gold nanosphere radius.}
\end{figure}

To quantify the size of the water layer that contributes to the photoacoustic generation, the following approach was implemented. The photoacoustic source term in Eq.\ref{eq:SolidPAEq} may be straightforwardly turned off in the simulations by forcing $\beta^f$ to zero at any desired locations. Several simulations were therefore run by limiting the extent of the photoacoustic source term to distances $r\in [R_s;R_s+\rho_{source}]$, with $\rho_{source}$ varied from 0 to $+\infty$. In practice, $\rho_{source}$ was varied up to a maximum value large enough so the photoacoustic signal did not differ significantly from its asymptotic value, corresponding to the case where all source points in water are active. The extent of the generating layer in water was then defined by the value $\rho_{layer}=\rho_{source}$ for which the amplitude of the photoacoustic signal reached $80\ \%$ of the amplitude of the asymptotic signal. This procedure was reiterated for different values of the laser pulse duration and the nanosphere radius, to compute the values of $\rho_{layer}(\tau_p, R_s)$ plotted on Fig.\ref{fig:Shell_dependence}.
Fig.\ref{fig:Shell_dependence} shows that the size of the contributing layer is in first approximation independent of the size of the sphere, and that it scales with the pulse duration approximately as $\rho_{layer}(\tau_p)\sim \sqrt{\tau_p}$. This scaling law suggests that the extent of the generating layer is dictated mostly by the diffusion of heat in water, regardless of the nanosphere diameter. As a consequence, each gold nanosphere may be considered as a nanometric absorber which thermally probes its environment within a spatial range driven by the laser pulse duration (longer than nanosecond). As an order of magnitude, a pulse duration $\tau_p = 5\ \mathrm{ns}$ yields $\rho_{layer} \sim 30\ \mathrm{nm}$.

\subsubsection{Comparison with the photoacoustic point-absorber model}\label{subsub:ComparisonPointSource}

It was shown above in Section \ref{subsub:OriginPA} that for a nanometric sphere illuminated with a nanosecond pulse, the photoacoustic wave is mostly generated by the liquid surrounding the particle. Within this regime, the analytical model proposed by~\citet{diebold2001} for point-absorbers in the linear regime is therefore expected to predict reasonably well the amplitude and shape of photoacoustic waves generated by gold nanospheres. The objective of this section is to quantify the accuracy of this theoretical model by comparing its predictions to our simulations for finite-size absorbers. This comparison will be further developed in the next results section for the nonlinear regime. From the analytical expression given by Eq.~\ref{eq:AnalyticDieboldLinear}, the photoacoustic energy emitted by a point absorber is given by
\begin{equation} \label{eq:CalassoLinearEnergy}
\epsilon_{\mathrm{point,linear}}=E_{abs}^2\frac{\beta_0^2}{c_p^3}\frac{1}{4 \pi}\frac{1}{\tau_p^3}\int_{-\infty}^{+\infty}\left[\frac{\dd f}{\dd u}\right]^2\dd u
\end{equation} 

\begin{figure}[h!]
\includegraphics[scale=1]{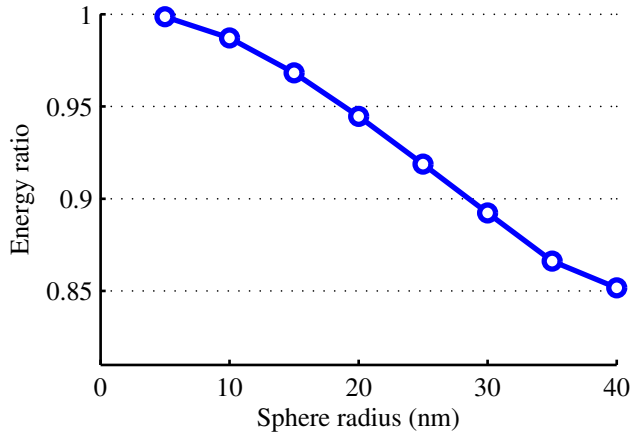} 
\caption{\label{fig:ComparisonDieboldSimu_lin} Ratio of the photoacoustic energy emitted from a gold nanosphere to that emitted from a point-absorber of identical absorption cross-section, as a function of the sphere dimensions, for a pulse duration $\tau_p=5 \mathrm{ns}$.}
\end{figure}
For gold nanospheres of different sizes, for a fixed pulse duration $\tau_p = 5~\mathrm{ns}$, we compared the emitted photoacoustic energy predicted by Eq.~\ref{eq:CalassoLinearEnergy} to that predicted for gold nanosphere in the linear regime, for equivalent absorption cross-sections. 
Fig. \ref{fig:ComparisonDieboldSimu_lin} shows a plot of $\alpha=\frac{\epsilon_{\mathrm{GNS, linear}}}{\epsilon_{\mathrm{point,linear}}}$ as a function of the nanosphere radius $R_s$. The predictions from the point absorber model and from our model turn out to be identical for vanishingly small diameters, as expected. Incidentally, this provides a validation of our numerical simulations in the linear regime. For finite sizes, the effect of the gold nanosphere, both as an acoustic scatterer and as a photoacoustic source (via $\beta^{s}$), is to marginally decrease the emitted acoustic energy compared to a point absorber of identical absorption cross-section. The effect is small, as expected from the very small ratio of the nanosphere diameter (typically tens of nm) to the acoustic wavelength (typically $20\ \mu\mathrm{m}$ for a 5-ns pulse). Therefore, in the nanosecond pulse regime, the point-absorber model proposed by~\citet{diebold2001} provides accurate quantitative predictions for the emission of photoacoustic waves by a gold nanosphere, with overestimation of the acoustic energy less than $10\ \%$ for sphere radii up to 30 nm.

\subsection{nonlinear regime}
In this section, we quantitatively investigate for a gold nanosphere the impact of the temperature dependence of the thermal expansion coefficient, that leads to the so called thermal photoacoustic nonlinearity. In a preamble section, we first discuss the consequences that can be derived from the analytical expression provided by \citet{diebold2001} within the point-absorber model. We then report quantitative predictions obtained for a gold nanosphere in the nonlinear regime.

\subsubsection{Predictions from the analytic point-absorber model}\label{subsub:PredictionFromNonLinearAnalyticPointModel}

\begin{figure}[!h]
\includegraphics[scale = 0.75]{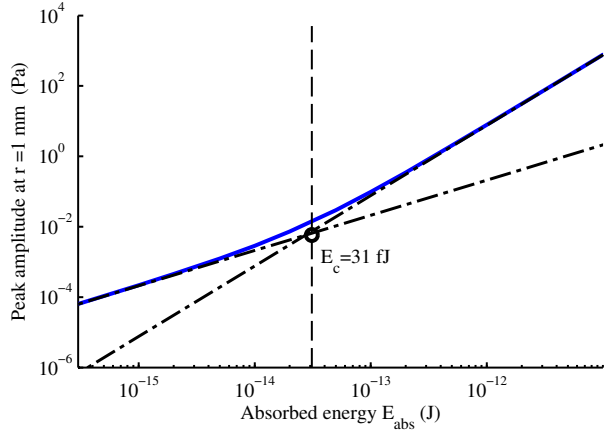} 
\caption{\label{fig:Dieboldprediction_log}  Peak photoacoustic amplitude (measured at r = 1 mm) as a function of absorbed energy, computed from the point-absorber model by \citet{diebold2001}, for an equilibrium temperature $T_{0}\ = \ 20\ ^\circ C$ and a pulse duration $\tau_p\ =\ 5\ \mathrm{ns}$.}
\end{figure}

\paragraph{Existence of a critical absorbed energy.}
\label{par:DefCriticalEnergy}

\begin{figure}[h!]
\includegraphics[scale=0.75]{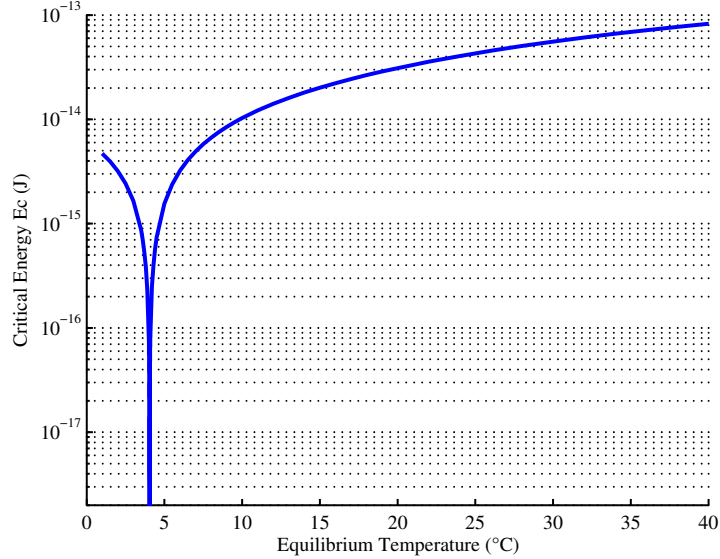} 
\caption{\label{fig:transition_figure} Critical energy $E_c$ as a function of the equilibrium temperature $T_{0}$}
\end{figure}
Eq.~\ref{eq:CalassoNonlinearPA} shows that for fixed physical constants, the relative contribution of the nonlinear term to the photoacoustic pressure wave only depends on the absorbed energy $E_{abs}$ and the pulse duration $\tau_p$. For a fixed pulse duration $\tau_p$=  5 ns, Fig.\ref{fig:Dieboldprediction_log} illustrates the change from the linear regime to the nonlinear regime as a function of the absorbed energy. We define the critical energy $E_c$ as the value of absorbed energy for which the peak amplitudes of the nonlinear and the linear contributions in Eq.~\ref{eq:CalassoNonlinearPA} are identical. With this definition, the nonlinear contribution is thus significant for $E_{abs}\gtrsim E_c$, and becomes predominant for $E_{abs}\gg E_c$. From Eq.~\ref{eq:CalassoNonlinearPA}, $E_c$ is given by 
\begin{equation}\label{eq:CriticalEnergy}
E_c=\frac{\mathrm{max}(\frac{\dd f}{\dd \hat{\tau}})}{\mathrm{max}(h)} \times \rho_0 c_p \frac{\beta_0}{\beta_1}(T_{0})\sqrt{\tau_p \chi}^3
\end{equation}
The numerical prefactor can be computed numerically from the function $h$ given by Eq.~(\ref{eq:h_function}) in Appendix~\ref{appendix:DerivationCalasso}, and yields
\begin{equation}\label{eq:CriticalEnergy}
E_c\simeq 20.2 \times \rho_0 c_p \frac{\beta_0}{\beta_1}(T_{0})\sqrt{\tau_p \chi}^3
\end{equation}

As indicated on Fig.~\ref{fig:Dieboldprediction_log}, $E_c=$ 31 fJ for $\tau_p$= 5 ns and $T_{0}\ =\ 20\ ^\circ C$.
By coupling Eq.~\ref{eq:CriticalEnergy} with the evolution of $\beta(T)$ in water shown in Fig.~\ref{fig:Beta}, the temperature-dependence of the critical energy was computed and plotted on Fig.~\ref{fig:transition_figure}. As $\beta$ vanishes around $4^\circ C$, so does $E_c$, i.e. only nonlinear generation is predicted for any absorbed energy at that temperature. More importantly, the curve in Fig.~\ref{fig:transition_figure} indicates that the critical energy is highly sensitive to the equilibrium temperature in a narrow range of a few degrees around $4^\circ C$. On the contrary, the critical energy is only weakly dependent on temperature at physiological temperatures (variation less than an order of magnitude for several tens of degrees).

Interestingly, the critical energy given by the point-absorber model scales as the volume around the absorber over which heat diffuses during $\tau_p$. The criteria given by equation (3) by~\citet{egerev2008} for a spherical particle of radius $R_s$ can be restated as a critical energy with a form similar to that of the point-absorber model:
\begin{equation}
\label{eq:Ec_egerev}
E_c\propto \rho_0 c_p \frac{\beta_0}{\beta_1}(T_{0})R_s^3  
\end{equation}
However, the volume term found by~\citet{egerev2008} is the volume of the sphere, whereas it is given from~\citet{diebold2001} by the volume of heat diffusion in water during the illumination pulse. These two predictions are incompatible: the critical energy given by Eq.~\ref{eq:Ec_egerev} goes to zero for vanishingly small spheres ($R_s \to 0$), whereas the point-absorber model  predicts a finite critical energy. Nevertheless, the two different expressions of the critical energy have in common to provide as a strong physical insight that the critical energy scales as some volume which origin remains to be determined. This will be further discussed based on our results for gold nanospheres in section~\ref{subsubsec:optimal size for nonlinear generation}. 

\begin{figure}[h!]
\includegraphics[scale=0.75]{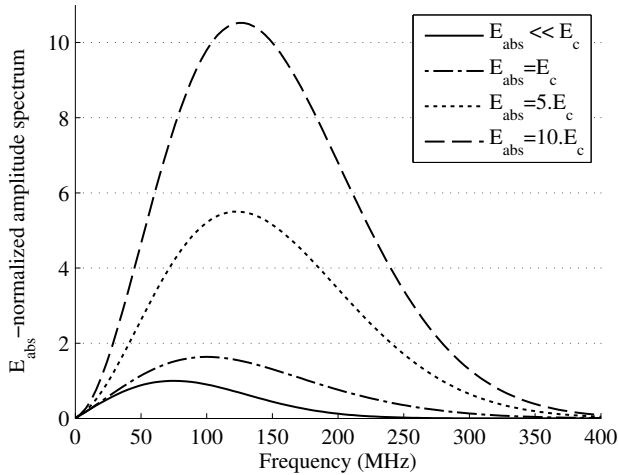} 
\caption{\label{fig:SpectraDiebold} Photoacoustic amplitude spectra as a function of the absorbed energy, for $\tau_p\ =\ 5 \ \mathrm{ns}$ at $T_{0}\ =\ 20\ ^\circ \mathrm{C}$. Each spectrum was normalized by the corresponding absorbed energy $E_{abs}$ in order to illustrate the nonlinear dependence on $E_{abs}$.}
\end{figure}

\paragraph{Frequency considerations.} The difference in temporal shapes for the linear and the nonlinear contributions was pointed out by \citet{diebold2001}. Here, we investigate the consequences of these different temporal shapes  (bipolar for the linear term, tripolar for the nonlinear term) in the frequency domain. For $\tau_p\ =\ 5 \ \mathrm{ns}$, Fig.~\ref{fig:SpectraDiebold} shows the frequency spectra of the photoacoustic signal as a function of the absorbed energy, with each spectrum normalized by the absorbed energy. 
Fig.~\ref{fig:SpectraDiebold} shows that the spectrum amplitude varies nonlinearly with the absorbed energy as expected from  Eq.~\ref{eq:CalassoNonlinearPA} in the time domain. Importantly, it also indicates that the frequency content is shifted towards high frequencies in the nonlinear regime. Therefore, the photoacoustic nonlinearity predicted by \citet{diebold2001} only manifests itself for sufficiently high frequency. This is coherent with the fact that the critical energy decreases with decreasing pulse duration (increasing centre frequency), as indicated by Eq.~\ref{eq:CriticalEnergy}. As a major consequence from the experimental point of view, Fig.~\ref{fig:SpectraDiebold} shows that even when the nonlinearity is predominant when considered over the full bandwidth, it remains minor for frequencies below 10 MHz even for $E_{abs}$ as high as $10\times E_c$.

\subsubsection{Temperature rise in a gold nanosphere}

\begin{figure}[h!]
\includegraphics[scale=1]{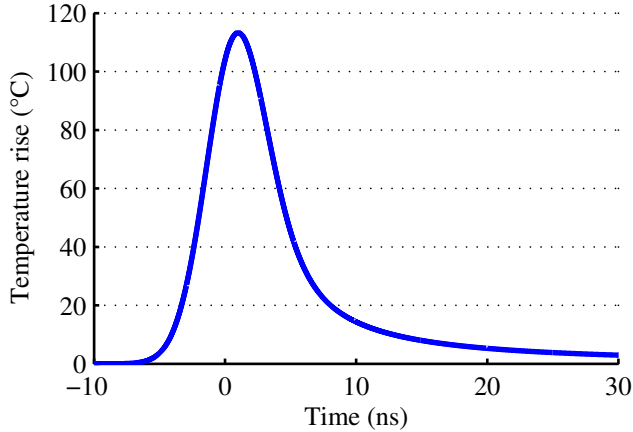} 
\caption{\label{fig:TemperatureRise} Temperature rise inside a 20-nm radius gold nanosphere illuminated with a fluence $\Phi_0=5\mathrm{mJ.cm^{-2}}$ ($E_{abs}\sim 165 \ \mathrm{fJ}$) and $\tau_p\ =\ 5\ \mathrm{ns}$.}
\end{figure}

\begin{figure}
\includegraphics[scale=1]{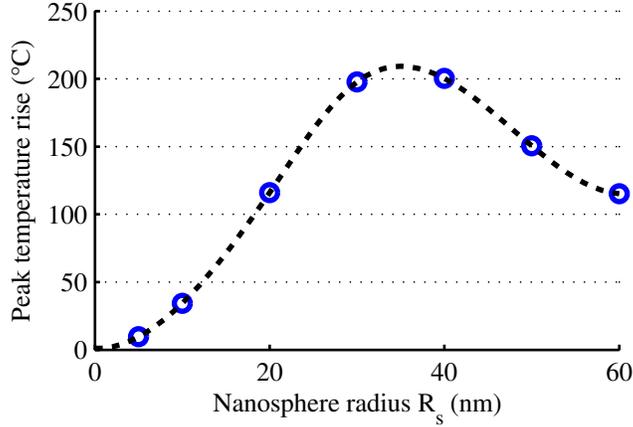} 
\caption{\label{fig:PeakTemperatureRise} Peak temperature rise as a function of gold nanosphere radius, for a fluence $\Phi_0=5\mathrm{mJ.cm^{-2}}$ ($E_{abs}\sim 165 \ \mathrm{fJ}$) and $\tau_p\ =\ 5\ \mathrm{ns}$. }
\end{figure}

It was shown in Section~\ref{subsub:ComparisonPointSource} in the limits of the linear model that the photoacoustic wave generated by a gold nanosphere is very close to that generated by a point absorber of identical absorption cross-section, and therefore depends only on the absorbed energy $E_{abs}$, with no dependency on thermal properties. In the case of thermal nonlinearity (caused by the temperature-dependence of the thermal expansion coefficient), it is the temperature field that drives the effective value of $\beta(T)$. As a consequence, because the temperature fields are different for a point absorber and a gold nanosphere with the same absorption cross-section, one expects the thermal photoacoustic nonlinearity to be dependent on the size of the absorber. As opposed to the point-absorber model, all the values of the temperature field are finite when finite-size absorbers such as gold nanospheres are considered. Moreover, for a given incident fluence, the temperature rise in a nanosphere is highly dependent on its size, both through the size-dependence of the thermal diffusion (the thermal Green's function given in Appendix~\ref{appendix:GreenTh} depends on the particle radius) and the absorption cross-section (see table~\ref{tab:Cross-Sections}). Fig.~\ref{fig:TemperatureRise} illustrates the temperature rise in a 40-nm diameter gold nanosphere illuminated with a fluence $\Phi_0=5\mathrm{mJ.cm^{-2}}$ ($E_{abs}\sim 165 \ \mathrm{fJ}$) and $\tau_p\ =\ 5\ \mathrm{ns}$. This plot shows that for a fluence value typical of those used for biomedical applications, the temperature rise in a 40-nm diameter gold nanosphere is  significantly larger than the equilibrium temperature. In this case, the photoacoustic nonlinearity is likely to become significant, as demonstrated further below. For a given illumination fluence, Fig.~\ref{fig:PeakTemperatureRise} shows that the peak temperature rise depends on the sphere radius and turns out to be maximum for a radius value around 35 nm,  reflecting the dependence on radius via both thermal diffusion and the size-dependence of the absorption cross-section.

\subsubsection{Gold nanosphere vs. point absorber} 

\begin{figure}[h!]
\includegraphics[scale=1]{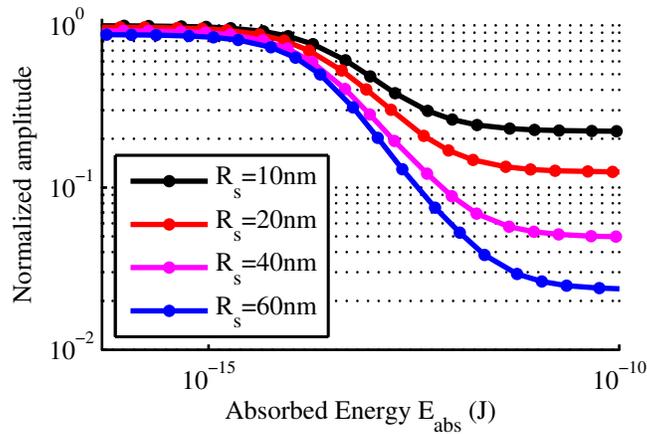} 
\caption{\label{fig:ComparisonDieboldvsGNS} Photoacoustic amplitude from a gold nanosphere, normalized to that of the point-absorber model, as a function of nanosphere diameter. Parameters: $T_{0}=20\ ^\circ  \mathrm{C}$, $\tau_p\ =\ 5\ \mathrm{ns}$}
\end{figure}

Simulations were run to predict the photoacoustic signals generated from gold nanospheres of various diameters illuminated with 5-ns pulses of various fluences, in order to compare the photoacoustic amplitude to that  predicted from the point-absorber model~\citep{diebold2001} with matched absorption cross-sections. Fig.~\ref{fig:ComparisonDieboldvsGNS} shows a plot of the ratio of the photoacoustic peak amplitude predicted for the nanosphere to that predicted for the point-absorber as a function of the absorbed energy. Figure ~\ref{fig:ComparisonDieboldvsGNS} confirms that in the linear regime, i.e for low enough absorbed energy (or equivalently $E_{abs}\ll E_c$), a gold nanosphere may be considered as a point-absorber, i.e. the predictions from the corresponding point-absorber model are accurate. However, in the nonlinear regime, the point-absorber model significantly overestimates the amplitude of the photoacoustic  signals. In other words, the critical energy for a gold nanosphere  is significantly higher than that predicted by the  point-absorber model.  For the gold nanosphere, in order to define the critical energy in agreement with the definition for the point absorber (see sec~\ref{par:DefCriticalEnergy}), the nonlinear contribution was defined by the difference between the total signal predicted in the nonlinear regime and the signal predicted in the linear regime only (by keeping $\beta$ constant).
\begin{figure}[h!]
\includegraphics[scale = 1]{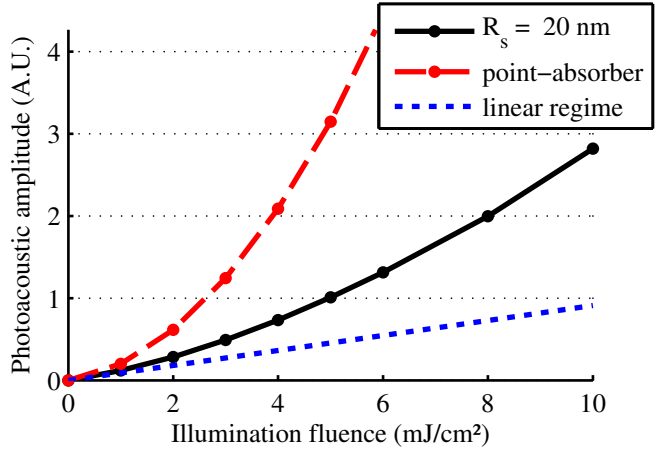} 
\caption{\label{fig:ComparisonDieboldvsGNS20nm} Photoacoustic peak amplitude for a 20-nm radius gold nanosphere and a point-absorber of identical absorption cross-section. $T_{0}=20\ ^\circ  \mathrm{C}$ and $\tau_p\ =\ 5\ \mathrm{ns}$}
\end{figure}
As an illustration, the photoacoustic amplitude predicted for a 20-nm radius gold nanosphere at equilibrium temperature $T_{0}=20\ ^\circ  \mathrm{C}$, illuminated with a 5-ns pulse of fluence $\Phi_0=5\mathrm{mJ.cm^{-2}}$ ($E_{abs}\sim 165 \ \mathrm{fJ}$), is about three times lower than that predicted by the point-absorber model, as illustrated in Fig.~\ref{fig:ComparisonDieboldvsGNS20nm}, and the critical energy is 216 fJ whereas the point-absorber model predicts a value of 31 fJ. On the other hand, the frequency features discussed in the previous section for the point-absorber model remain strictly identical for a gold nanosphere, i.e. the nonlinearity is observed only at high enough frequencies. Fig.~\ref{fig:SpectraDiebold} for the point-absorber remains valid for a gold nanosphere, provided that the appropriate value of critical energy is used. In conclusion to this comparison between our predictions for gold nanospheres and predictions for the point-absorber model, the finite size of the gold nanosphere must be taken into account to obtain accurate quantitative predictions regarding the occurence of thermal-based photoacoustic nonlinearities.

\begin{table*}[h!]
\caption{Values of the critical energy $E_c$, critical fluence $\Phi_c$ and maximum temperature rise $\Delta T_{max}$ for gold nanospheres as a function of radius $R_s$ and equilibrium temperature $T_{0}$, with $\tau_p\  =\  5 \ \mathrm{ns}$.}
\begin{tabular}{|c|>{\centering}p{3 cm}|c*{5}{|c}|} 
        \hline     
$T_0(^\circ C) \backslash R_s$ (nm)& & 10 & 20 & 30 & 40 & 50 & 60       \\
    \hline     
    \multirow{3}{*}{5} & $E_c^{R_s} (\mathrm{fJ})$ 
    & 9.4 & 19 & 38 & 66 & 110 & 170\\
    & $\Phi_c^{R_s} (\mathrm{mJ}/\mathrm{cm}^2)$ 
    & 2.5 & 0.59 & 0.35 & 0.36 & 0.51 & 0.74\\
    							& $\Delta T_{max} (^\circ C)$ 
    & 17.4 & 13.7 & 13.8 & 14.5 & 15.5 & 17.1\\
    \hline     
    \multirow{3}{*}{10} & $E_c^{R_s} (\mathrm{fJ})$ 
    & 52 & 93 & 150 & 230 & 340 & 490\\
    & $\Phi_c^{R_s} (\mathrm{mJ}/\mathrm{cm}^2)$ 
    & 14 & 2.8 & 1.4 & 1.3 & 1.6 & 2.1\\
    							& $\Delta T_{max} (^\circ C)$ 
    & 96 & 65 & 55 & 51 & 49 & 49\\
    \hline     
    \multirow{3}{*}{20} & $E_c^{R_s} (\mathrm{fJ})$ 
    & 120 & 220 & 350 & 520 & 740 & 1040\\
    & $\Phi_c^{R_s} (\mathrm{mJ}/\mathrm{cm}^2)$ 
    & 33 & 6.5 & 3.2 & 2.9 & 3.6 & 4.5\\
    							& $\Delta T_{max} (^\circ C)$ 
    & 230 & 150 & 126 & 114 & 107 & 104\\
    \hline     
    \multirow{3}{*}{40} & $E_c^{R_s} (\mathrm{fJ})$ 
    & 230 & 400 & 630 & 940 & 1350 & 1850\\
    & $\Phi_c^{R_s} (\mathrm{mJ}/\mathrm{cm}^2)$ 
    & 62 & 12 & 5.8 & 5.1 & 6.4 & 8.1\\
    							& $\Delta T_{max} (^\circ C)$ 
    & 420 & 280 & 230 & 210 & 190 & 190\\
    \hline     
    
\end{tabular}
\label{tab:BigTable_Ec_Phic_Tmax}
\end{table*}

\subsubsection{Optimal size for nonlinear generation} 
\label{subsubsec:optimal size for nonlinear generation}

In order to quantitatively predict the occurence of nonlinear photoacoustic generation as a function of  parameters that are controllable experimentally,  simulations were run for gold nanospheres with different radii and equilibrium temperatures. Table.~\ref{tab:BigTable_Ec_Phic_Tmax} reports the corresponding results as the value of the critical energy $E_c$, critical fluence $\Phi_c=E_c/\sigma_a$ and the peak temperature in the sphere. In particular, these results show that for any equilibrium temperature, there is an optimal sphere radius around 40 nm for which the critical fluence is minimal (as a function of size). In other words, our model predicts that at a given illumination fluence, the thermal-based photoacoustic nonlinearity is maximised for a sphere radius around 40 nm. We note that the value of the gold nanosphere radius that maximizes the photoacoustic nonlinearity (for a given illumination fluence) is close to the one that maximizes the peak temperature rise in the nanosphere (see Fig.~\ref{fig:PeakTemperatureRise}): because our model assumes the continuity of temperature across the gold/water interface, the peak temperature in water is also maximized for a radius of 40 nm, which is expected to maximize thermal nonlinearity.
\begin{figure}[h!]
\includegraphics[scale=1]{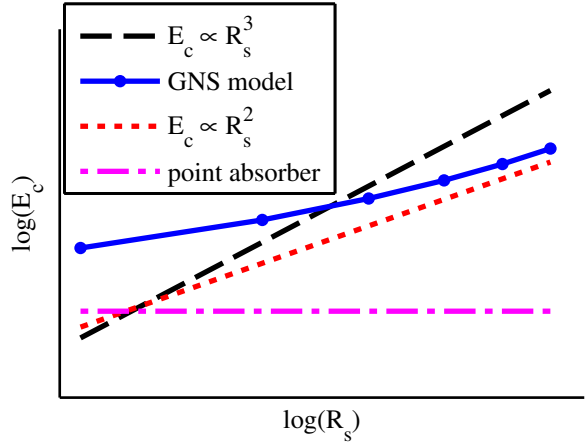}
\caption{\label{fig:Ec_f_R} Critical energy as a function of nanosphere radius (logarithmic axis).
}
\end{figure}
The values of critical energy given in Table.~\ref{tab:BigTable_Ec_Phic_Tmax} can be further analyzed as a function of the particle size, for a given equilibrium temperature, in order to assess the relevance of the scaling law predicted by the point-absorber model ($E_c$ independent of $R_s$, see Eq.~\ref{eq:CriticalEnergy}) and the model by~\citet{egerev2008} ($E_c \propto R_s^3$, see Eq.~\ref{eq:Ec_egerev}). For $T_0=20\ ^{\circ}C$, Fig~\ref{fig:Ec_f_R} shows a plot of the values of the critical energy predicted by our model, by the point-absorber model, and for $E_c \propto R_s^3$ and $E_c \propto R_s^2$. This plot is compatible with the hypothesis that the critical energy tends towards the constant value of the point-absorber for small spheres, and scales as the surface $R_s^2$ for large spheres. For large spheres, it is indeed expected that the volume of heat diffusion scales as the surface of the sphere times the radial distance traveled by heat (only dependent on the pulse duration and the diffusivity in water). Beyond providing quantitative values for the critical energy (or equivalently the critical fluence via the absorption cross-section), our model therefore also provides some physical insight into the origin of the critical energy.

\subsubsection{Influence of the equilibrium temperature}

\begin{figure}[h!]
\includegraphics[scale=1]{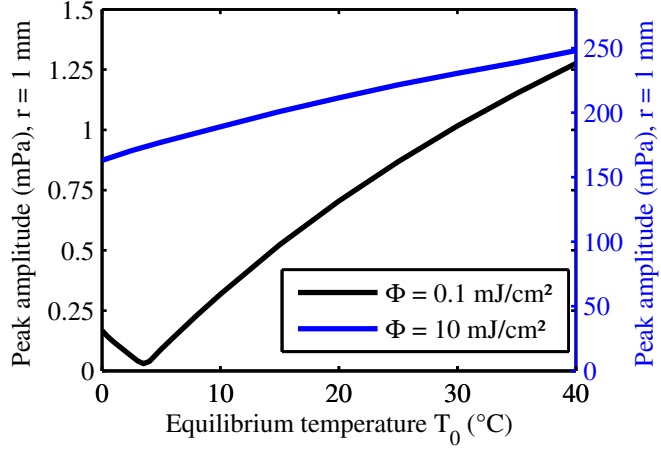}
\caption{\label{fig:effetTeq}Peak photoacoustic amplitude emitted by a 40-nm diameter gold nanosphere, as a function of equilibrium temperature, for two different fluence $\Phi$. $\Phi=0.1\ \mathrm{mJ/cm^2}$ corresponds to the linear regime, whereas significant nonlinearities occur at $\Phi=10\ \mathrm{mJ/cm^2}$.
}

\end{figure}

In this section, we study the influence of the equilibrium temperature on the photoacoustic amplitude, for a fixed illumination fluence. Fig.~\ref{fig:effetTeq} shows the results obtained in the linear regime (lower black curve) and in the nonlinear regime (upper blue curve). In the linear regime, the photoacoustic amplitude reflects the temperature dependence of the thermal expansion coefficient $\beta_0(T_0)$, as expected from Eq.~\ref{eq:PAEqHomogeneous}. In the nonlinear regime, at a higher illumination fluence, the temperature-dependence of the signal is much less significant, and essentially reflects the effective value of the thermal expansion coefficient, different from the value at equilibrium because of the significant temperature rise around the particle. As a consequence, the evolution of the peak photoacoustic amplitude as a function of equilibrium temperature reflects temperature rises at the nanoscale. In particular, the strong signal observed at $T_0 = 4 \ ^\circ C$ for which $\beta_0=0$ is purely nonlinear in nature as the linear contribution vanishes at this temperature. This has first been observed experimentally by~\citet{hunter1981} in protons experiments and has been more recently reported with gold nanoparticles by~\citet{fukasawa2014} and~\citet{simandoux2014}. Recent experimental measurements of the photoacoustic amplitudes as a function of equilibrium temperature with 40-nm diameter gold nanospheres by~\citet{simandoux2014} are in good agreement with the predictions given in Fig.~\ref{fig:effetTeq}.

An important consequence of the results shown in Fig.~\ref{fig:effetTeq} is that when thermal nonlinearity takes place, the amplitude of photoacoustic waves cannot provide a measurement of the equilibrium temperature through the knowledge or calibration of $\beta_0(T_0)$. While photoacoustic measurements can provide a measurement of the equilibrium temperature in the linear regime, as was originally proposed by~\citet{larina2005} and further investigated in subsequent works (see ~\citep{shah2008,Petrova2013,Gao2013} for instance), this approach is expected to not work any longer with nanoparticles illuminated with high enough fluences.

\subsection{Discussion}
\label{subsec:Discussion}

The photoacoustic amplitude predicted by the point absorber model (see Eq.~\ref{eq:AnalyticDieboldLinear}) is independent of thermal diffusion in water, and only depends on the absorbed energy. Therefore our results suggest that when gold nanospheres behave as point-absorbers in the \textit{linear} regime, the photoacoustic amplitude is insensitive to heat transfer from gold to water, although the temperature field does depend on the thermal diffusivity of water (see Appendix~\ref{appendix:GreenTh}). On the contrary, \textit{nonlinear} photoacoustic signals are strongly dependent on the thermal diffusivity of water (as predicted by the point-absorber model in water by Eq.~\ref{eq:CalassoNonlinearPA}). This suggest that the occurence of thermal nonlinearity is required to probe thermal diffusion properties at the spatial and temporal nanoscale. In the model described here, the temperature was assumed to be continuous across the gold/water interface, whereas it is known that an interfacial thermal resistivity exists and may have significant impact at the nanoscale~\citep{alper2010,schmidt2008,wilson2002}. To check the importance of this effect on our predictions, we solved the thermal problem by taking into account an interfacial thermal resistance at the gold-water interface. The temperature field for a delta impulse illumination (Green's function) was found to strongly depend on the thermal resistance, in agreement with earlier work~\citep{juve2009}. However, when convolved with a 5-ns pulse illumination, the temperature field for a interfacial resistance of $10^{-8}\ \mathrm{m^2.K.W^{-1}}$, typical of the gold-water interface, was nearly identical to the case with no thermal resistance (less than a few $\%$ relative difference). The results presented in this work with no interfacial thermal resistance therefore remains valid in the few-nanosecond regime. On the other hand, the presence of coating, such as silica coatings or PEG (poly-ethylene glycol) coatings, may have a significant effect, and will be investigated in a future work.

Importantly, our work describes the photoacoustic generation by a \textit{single} gold nanosphere. When collections of nanoparticles such as encountered in suspension are considered, the emitted photoacoustic waves arise from the sum of each contribution from individual particles. In this case, the frequency content of the resulting wave is dictated by the spatial distribution of the collection of nanoparticles, which acts as a low-pass filter. Because the thermal nonlinearity was shown to be significant only at high frequencies, it is expected that it may not be observable at low frequencies. In a recent experimental study~\citep{simandoux2014} with a detection frequency of 20 MHz and 40-nm diameter gold nanospheres suspended in $100\ \mathrm{\mu m}$ diameter tube, a linear relationship was observed between the peak-to-peak amplitude and the fluence, up to a fluence of $\Phi=7\ \mathrm{mJ/cm^2}$, yet close to the corresponding critical fluence predicted here $\Phi_c = 6.5\ \mathrm{mJ/cm^2}$. However, as discussed above, the critical fluences predicted here are only valid for a single nanoparticle, and the prediction for ensembles of nanoparticles, beyond the scope of this work, requires taking into account the nanoparticles spatial distribution. 

It also has to be kept in mind that our model assumes that there is no limitation on the peak temperature in the gold nanosphere and in water. In practice, the model becomes of course irrelevant if the predicted temperatures would lead to phase transitions in either gold or water. In addition, the thermal model used in our study is strictly restricted to the long-pulse illumination regime (pulses typically longer than a few picosecond), and the corresponding modeling cannot be applied to the sub-picosecond illumination regime which usually requires either a two-temperature model~\cite{eesley1983} or nonthermal distributions~\citep{tas1994,groeneveld1995} for the electrons and phonons in nanoparticles.

\section{Summary and conclusion}
In summary, we theoretically investigated the photoacoustic generation by a gold nanosphere in water in the thermoelastic regime. Photoacoustic signals were predicted numerically based on the successive resolution of a thermal diffusion problem and a thermoelastic problem, taking into account the finite size of the gold nanosphere, thermoelastic and elastic properties of both water and gold, and the temperature dependence of the thermal expansion coefficient of water. For sufficiently high illumination fluences, this temperature-dependence yields a nonlinear relationship between the photoacoustic amplitude and the fluence. For nanosecond pulses in the linear regime, we showed that more than $90\ \%$ of the emitted photoacoustic energy is generated in water, and the thickness of the generating layer around the particle scales close to the square root of the pulse duration. In the linear regime, we showed that the point-absorber model introduced by~\citet{diebold2001} accurately predicts the amplitude of the photoacoustic waves generated by gold nanospheres for diameters up to several tens of nanometers. However, the point-absorber model significantly overestimates the amplitude of photoacoustic waves generated by gold nanospheres in the nonlinear regime. Our model for finite-size particles provided quantitative estimates of the critical energy, defined as the absorbed energy required such that the nonlinear contribution is equal to that of the linear contribution. Our results suggest that the critical energy scales as the volume of water over which heat diffuses during the illumination pulse. A frequency analysis of the nonlinear signals indicated that the thermal nonlinearity from a gold nanosphere is more pronounced at high frequencies dictated by the pulse duration. Finally, we show that the relationship between the photoacoustic amplitude and the equilibrium temperature at sufficiently high fluence reflects the thermal diffusion at the nanoscale around the gold nanosphere. Although our model was limited to the case of a bare nanoparticle, the approach used in this work can be extended to the more general case of a coated nanoparticle, provided that the corresponding temperature field can be predicted, either by analytical or numerical means (such as given by~\citet{baffou2011}). To limit the computational costs, and for sake of simplicity, this work was limited to nanospheres with central symmetry. However, the same methodology could apply in two dimensions for axisymmetric nanoparticles such a gold nanorods at the cost of more intensive computations.

\section{Acknowledgements}
This work was funded by the French Centre National de la
Recherche Scientifique, the Plan Cancer 2009-2013 (Action
1.1, Gold Fever), and the Agence Nationale de la Recherche
(Golden Eye, ANR-10-INTB-1003, and the LABEX WIFI
within the French Program 'Investments for the Future', ANR-
10-LABX-24 and ANR-10-IDEX-0001-02 PSL*).

\begin{appendix}

\section{Nonlinear thermoelastic regime}
\label{appendix:NonlinearRegime}

\subsection{Derivation of Eq.~\ref{eq:NonLinearPAfluid}}
\label{appendix:DerivationEquation}

We consider the linear regime of acoustic propagation, \textit{i.e.} small Mach number ($\frac{\|\mathbf{v}\|}{c_s}\ll 1$) and small density changes ($\frac{\rho-\rho_0}{\rho_0}\ll 1$). In particular, the following approximation holds: $\frac{\delta X}{\delta t}=\frac{\partial X}{\partial t}+(\mathbf{v}.\nabla)X\simeq \frac{\partial X}{\partial t}$. The three fundamental equations of linear acoustics can be written as
\begin{align}
\frac{\partial\rho}{\partial t}&=-\rho_0\ \mathrm{div}(\mathbf{v}) \label{eq:ContinuityEq}\\
\rho_0\frac{\partial \mathbf{v}}{\partial t}&= -\nabla p \label{eq:Euler}\\
\rho T\frac{\partial s}{\partial t}&=\kappa\Delta T +P_v \label{eq:EntropicDiffusion}
\end{align}
where $p(\mathbf{r},t)$ is the acoustic pressure, $\mathbf{v}(\mathbf{r},t)$ is the acoustic displacement velocity, $s(\mathbf{r},t)$ is the specific entropy and $T(\mathbf{r},t)$ is the temperature. The thermal conduction coefficient $\kappa$ is assumed to be constant. Moreover, the differentials of the state functions $\rho=\rho(p,T)$ et $s=s(T,p)$ can be written as \citep{morse1986}
\begin{align}
\delta \rho &=\frac{\gamma}{c_s^2}[\delta p   -\beta \delta T]\label{eq:evolution drho}\\
\delta s&=\frac{c_p}{T}[\delta T + \frac{\gamma-1}{\gamma \beta}\delta p ]\label{eq:evolution ds}
\end{align}
where the thermodynamical coefficients are defined as
\begin{align*}
c_s^2=\left(\frac{\partial p}{\partial \rho}\right)_s \\
c_p=T \left(\frac{\partial s}{\partial T}\right)_{p} \\
c_v=T\left(\frac{\partial s}{\partial T}\right)_{\rho} \\
\gamma=\frac{c_p}{c_v}\\
\beta=-\frac{1}{\rho}\left(\frac{\partial \rho}{\partial T}\right)_{p}
\end{align*}
with $c_s^2$ the isentropic sound velocity, $c_p$ the specific heat capacity at constant pressure,  $c_v$ the specific heat capacity at constant volume and $\beta$ the thermal expansion coefficient. Eqs.~\ref{eq:evolution drho} and \ref{eq:evolution ds} may be written as
\begin{align}
\frac{\partial\rho}{\partial t}&=\frac{\gamma}{c_s^2}[\frac{\partial p}{\partial t}   -\beta \frac{\partial T}{\partial t}]\label{eq:evolution temporelle rho}\\
\frac{\partial s}{\partial t}&=\frac{c_p}{T}[ \frac{\partial T}{\partial t} + \frac{\gamma-1}{\gamma \beta}\frac{\partial p}{\partial t}]
\label{eq:evolution temporelle s}
\end{align}
where all the thermodynamical coefficients may vary with properties such as temperature. In the following, we consider that only the variations of $\beta(T)$ with temperature are significant, and assume all other thermodynamical properties to be constant.
By coupling the first two fundamental equations \ref{eq:ContinuityEq} and \ref{eq:Euler}, one gets 
$\frac{\partial^2 \rho}{\partial t^2}-\Delta p=0$, which can be transformed into the following equation by use of Eq.~\ref{eq:evolution temporelle rho}
\begin{equation}
\frac{\gamma}{c_s^2}\frac{\partial^2 p}{\partial t^2}-\Delta p =\frac{\rho_0 }{c_p}\frac{\partial }{\partial t}
\left(\beta(T) T \frac{\partial s}{\partial t} \right)
\end{equation}
By use of Eq.~(\ref{eq:evolution temporelle s}) into the evolution equation~(\ref{eq:evolution ds}) for the entropy, one obtains the following equation for the temperature field:
\begin{equation}
\rho_0 c_v \frac{\partial T}{\partial t}=\kappa\Delta T +P_v +\rho_0 c_v \frac{\gamma -1}{\gamma \beta}\frac{\partial p}{\partial t} \label{eq:équation diffusion température avec rho}
\end{equation}
Under the assumption, valid for liquids, that $\gamma \sim 1$, one eventually obtains Eqs.~(\ref{eq:ThermalEqHomogeneous}) and (\ref{eq:CalassoNonlinearPA}) given in section \ref{sub:PhysicalModel}:
\begin{align}
\rho_0 c_v \frac{\partial T}{\partial t}&=\kappa\Delta T +P_v\label{eq:équation diffusion température}\\
\frac{1}{c_s^2}\frac{\partial^2 p}{\partial t^2}-\Delta p &=\rho _0\frac{\partial }{\partial t}
\left(\beta(T)\frac{\partial T}{\partial t} \right)\label{eq:équation onde pression} 
\end{align}
Eq.~(\ref{eq:équation onde pression}) may also obtained from Eqs.~(\ref{eq:SolidPAEq}), by considering a liquid as a special type of solid for which $\mu=0$.

\subsection{Consequences from the form of Eq.~(\ref{eq:CalassoNonlinearPA})/(\ref{eq:équation onde pression})}
The photoacoustic wave equation Eq.~(\ref{eq:CalassoNonlinearPA})/(\ref{eq:équation onde pression}) used in this work and derived above, differs from the equation originally introduced by \citet{burmistrova1979} which is given only for small temperature variations around $T_0$ and reads:

\begin{equation}
\label{eq:Burmistrova} 
\frac{1}{c_s^2}\frac{\partial^2 p}{\partial t^2}-\Delta p =\rho _0\frac{\partial^2 }{\partial t^2}\left[ \beta(T) (T-T_0)\right]
\end{equation}

Eq.~(\ref{eq:Burmistrova}) corresponds to that given by \citet{gusev1993}, further used by \citet{diebold2001} for the analytic derivation corresponding to the point-absorber model. Although the two forms source term are very different in these two equations, it turns out as demonstrated below that the consequences on the solution are relatively minor when one consider only the first-order dependence of $\beta(T)$ with temperature, as $\beta(T)=\beta_0+\beta_1 (T-T_0)$ (with $\beta_0=\beta(T_0)$ and $\beta_1=\frac{\dd \beta}{\dd T}(T_0)$). By using this first-order expansion of $\beta(T)$ around $T_0$, our equation Eq.~(\ref{eq:CalassoNonlinearPA}) becomes
\begin{equation}
\frac{1}{c_s^2}\frac{\partial^2 p}{\partial t^2}-\Delta p =\rho_0\left[\left(\beta_0+\beta_1(T-T_0)\right)\frac{\partial^2 T}{\partial t^2}+\beta_1\left(\frac{\partial T}{\partial t}\right)^2\right]
\end{equation}
whereas the original equation used by \citet{burmistrova1979,gusev1993,diebold2001} becomes
\begin{equation}\label{eq:equation gusev beta linéarisé}
\frac{1}{c_s^2}\frac{\partial^2 p}{\partial t^2}-\Delta p =\rho_0\left[\left(\beta_0+2\beta_1(T-T_0)\right)\frac{\partial^2 T}{\partial t^2}+2\beta_1\left(\frac{\partial T}{\partial t}\right)^2\right]
\end{equation}

It therefore turns out that the only remaining difference is a factor 2 before $\beta_1$. In consequence, all the analytical calculations based on Eq.~(\ref{eq:Burmistrova}), as given by  \citet{burmistrova1979,gusev1993,diebold2001}, may be used straightforwardly by changing $\beta_1$ to $\beta_1/2$.

\subsection{Modified expression for the point-absorber model}
\label{appendix:DerivationCalasso}
Three modifications are needed to go from the original equation (25) given by \citet{diebold2001} to our Eq.~(\ref{eq:CalassoNonlinearPA}):
\begin{enumerate}
\item \citet{diebold2001} erroneously used a term $e^{r^2/2 \chi \xi }$ in their Eq. (19), instead of $e^{r^2/4 \chi \xi }$, as straightforward from their Eq. (6). As can be derived from simple calculations, the only consequence of this minor error is that equation (25) by \citet{diebold2001} has to be multiplied by a factor $2^{3/2}$.
\item In \citet{diebold2001}, the derivation of the nonlinear contribution to the photoacoustic signal by a point-absorber was based on Eq.~(\ref{eq:Burmistrova}) (or stricly speaking its equivalent formulated as a displacement potential), under the assumption that $\beta(T)=\beta_0+\beta_1 (T-T_0)$. From the discussion above, we therefore had to change the expression given in \citet{diebold2001} by replacing $\beta_1$ to $\beta_1/2$ to take into account our equation \ref{eq:CalassoNonlinearPA}. (N.B. \citet{diebold2001} used the notation $(\beta1,\beta2)$ whereas we used $(\beta0,\beta1)$). In other words, Eq (25) in \citet{diebold2001} simply has to be multiplied by a factor 1/2 to take into account Eq.~(\ref{eq:CalassoNonlinearPA}), but all the analytical derivations remain valid except for this numerical prefactor.\\
 
Overall, to take into account the two prefactors above, Eq (25) in~\citet{diebold2001} has to be multiplied by a factor $1/2\times2^{3/2}=2^{1/2}$.
\item The pulse duration $\theta$ used by \citet{diebold2001} is related to the pulse duration $\tau_p$ defined here by $\tau_p=2\sqrt{\ln(2)}\ \theta$. Moreover, we use the dimensionless retarded time $\hat{\tau}=(t-r/c_s)/\tau_p$ whereas \citet{diebold2001} used $\hat{\tau}=(t-r/c_s)/\theta$. This change of variable further adds a numerical prefactor to the expression of the nonlinear term.
\end{enumerate}
Taking into account the three modifications above, the nonlinear contribution $p^{NL}$ from the point-absorber model can be written as
\begin{equation}
p^{NL}(\mathbf{r},t)=E_{abs}^2\beta_1\frac{1}{\rho_0 \chi^{3/2} c_p^2 \tau_p^{7/2}}\frac{1}{4 \pi r}h\left(\hat{\tau}=\frac{t-\frac{r}{c_s}}{\tau_p}\right)
\end{equation}
where $h(\hat{\tau})$ is given by
\begin{equation}
\label{eq:h_function}
h(\hat{\tau})=\frac{[\ln(2)]^{3/4}}{8\pi^2} \frac{\dd^2}{\dd \hat{\tau}^2}\left[\int_{0}^{\infty}\frac{\mathrm{erf}(\xi/\sqrt{2})}{\xi^{3/2}}e^{-2\left(\xi/2-2\sqrt{\ln(2))}\hat{\tau}\right)^2}\dd \xi \right]
\end{equation}

\section{Green's function of the thermal problem}\label{appendix:GreenTh}
The following expression of the Green's function of the thermal problem may be found in \citet{egerev2009} and written as
\begin{align}
&G_{\mathrm{th}}(\mathbf{r},t)=\frac{1}{\alpha_1-\alpha_2}\frac{R_s}{r}  \Biggl[ \nonumber \\
&+\alpha_1 \exp\left(\alpha_1(\frac{r}{R_s}-1)+\alpha^2_1\frac{t}{\tau_{th}}\right)
 \mathrm{erfc}\left(\frac{\frac{r}{R_s}-1}{2\sqrt{\frac{t}{\tau_{th}}}}+\alpha_1\sqrt{\frac{t}{\tau_{th}}}\right)\nonumber \\ 
& -\alpha_2 \exp\left(\alpha_2(\frac{r}{R_s}-1)+\alpha^2_2\frac{t}{\tau_{th}}\right)
 \mathrm{erfc}\left(\frac{\frac{r}{R_s}-1}{2\sqrt{\frac{t}{\tau_{th}}}}+\alpha_2\sqrt{\frac{t}{\tau_{th}}}\right) \Biggr]
\end{align}

where: 
\begin{equation*}
\tau_{th}=\frac{R_s^2}{\chi}, \ 
\alpha_\frac{1}{2}=\frac{\eta}{2}\left(1\pm\sqrt{1-\frac{4}{\eta}}\right), \ 
\eta=3\frac{\rho_0^w c_p^w}{\rho_0^{Au}c^{Au}_p}
\end{equation*}

\section{FDTD algorithm for the thermoelastic problem}\label{appendix:FDTD}

\begin{figure}[]
\includegraphics[width=9 cm]{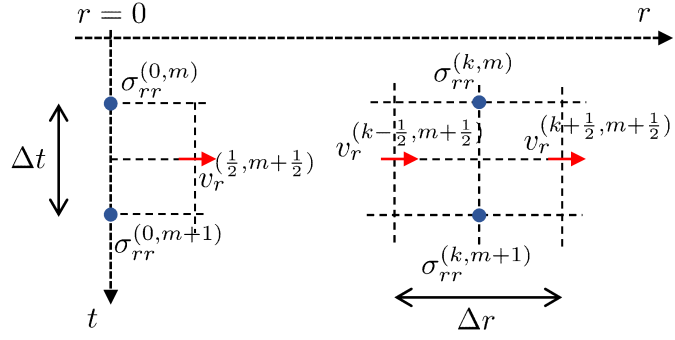} 
\caption{\label{fig:FDTD} Spatio-temporal mesh used in FDTD simulations. The stress field $\sigma_{rr}$ (blue points) and the radial velocity displacement (red arrows) $v$ are discretized over staggered grids, both in time and space \cite{virieux1986}. The grids are staggered such that pressure variables are defined on the central point $r=0$. Values of $\sigma_{\theta\theta}$ and $T$ are defined at the same positions as $\sigma_{rr}$.}
\end{figure}

Following the approach introduced for elastodynamics by~\citet{virieux1986}, Eqs.~\ref{eq:SolidPAEq} were discretized on staggered grids (see Fig.~\ref{fig:FDTD}) with the following centered finite-difference approximation:
\begin{equation}\label{eq: discrétisation définition base}
\frac{\partial f}{\partial a}(a_i)\approx \frac{f(a_i+\frac{\Delta a}{2})-f(a_i-\frac{\Delta a}{2})}{\Delta a}
\end{equation}
where $a$ may be either $r$ or $t$. Moreover, whenever needed at locations where they were undefined (either field variables or material properties), variables were approximated by their arithmetic mean with $f(a_i)=\frac{f(a_i+\frac{\Delta a}{2})-f(a_i-\frac{\Delta a}{2})}{2}$. All material properties were defined at locations ${k\Delta r}$.  The Eqs.~\ref{eq:SolidPAEq} accordingly read
\begin{subequations}
\begin{align}\label{eq: Equa PA solide vitesse discretisee}
&\frac{v_r(k+\frac{1}{2},m+\frac{1}{2})-v_r(k+\frac{1}{2},m-\frac{1}{2})}{\Delta t}= \frac{2}{\rho_0(k+1)+\rho_0(k)}\times \left[\right.\nonumber  \\
&\frac{\sigma_{rr}(k+1,m)- \sigma_{rr}(k,m)}{\Delta r} + \nonumber \\
& \frac{2}{\Delta r(k+\frac{1}{2})} \times \left.\left[\frac{\sigma_{rr}(k+1,m)+\sigma_{rr}(k,m)}{2}-\frac{\sigma_{\theta\theta}(k+1,m)+\sigma_{\theta\theta}(k,m)}{2}\right]\right] \\
\nonumber \\
&\frac{\sigma_{rr}(k,m+1)-\sigma_{rr}(k,m)}{\Delta t}= \nonumber \\
& \lambda(k)\frac{1}{\Delta r}\frac{1}{k^2}\left[\left(k+\frac{1}{2}\right)^2 v_r(k+\frac{1}{2},m+\frac{1}{2}) -\left(k-\frac{1}{2}\right)^2 
 v_r(k-\frac{1}{2},m+\frac{1}{2}) \right] + \nonumber\\
& 2\mu(k)\frac{1}{\Delta r} \left[v_r(k+\frac{1}{2},m+\frac{1}{2})- v_r(k-\frac{1}{2},m+\frac{1}{2}) \right] - \nonumber\\
& \left[\lambda(k)+\frac{2}{3}\mu(k)\right] \beta(T(k,m+\frac{1}{2})) \frac{T(k,m+1)-T(k,m)}{\Delta t}  \\
\nonumber \\
& \frac{\sigma_{\theta\theta}(k,m+1)-\sigma_{\theta\theta}(k,m)}{\Delta t}= \nonumber \\
&\lambda(k)\frac{1}{\Delta r}\frac{1}{k^2}\left[\left(k+\frac{1}{2}\right)^2 v_r(k+\frac{1}{2},m+\frac{1}{2}) -\left(k-\frac{1}{2}\right)^2 
v_r(k-\frac{1}{2},m+\frac{1}{2}) \right] + \nonumber\\ 
&2\mu(k)\frac{1}{\Delta r} \frac{1}{2}\left[\frac{v_r(k+\frac{1}{2},m+\frac{1}{2})}{k+\frac{1}{2}} + \frac{v_r(k-\frac{1}{2},m+\frac{1}{2})}{k-\frac{1}{2}}  \right]  - \nonumber \\
& \left[\lambda(k)+\frac{2}{3}\mu(k)\right] \beta(T(k,m+\frac{1}{2})) \frac{T(k,m+1)-T(k,m)}{\Delta t}
\end{align}
\label{eq:EqPAsolidFDTD}
\end{subequations}
where we used $\left(\frac{\partial}{\partial r}+\frac{2}{r}\right)v_r=\frac{1}{r^2}\frac{\partial r^2 v_r}{\partial r}$. At the central point, Eqs.~\ref{eq:EqPAsolidFDTD}b and~\ref{eq:EqPAsolidFDTD}c cannot be used because of the singularity in $1/r$. At $r=0$, we used $\frac{v_r}{r}\sim r \frac{\partial v_r}{\partial r}$ and $v_r(0)=0$ to obtain the following form 
\begin{subequations}
\begin{align} 
\frac{\partial \sigma_{rr}}{\partial t}(0,t)&= (3\lambda(0)+2\mu(0))\frac{\partial v_r}{\partial r}(0,t)  - \left( \lambda(0)+\frac{2}{3}\mu(0)\right)\beta(T) \frac{\partial T}{\partial t}(0,t) \\
\frac{\partial \sigma_{\theta,\theta}}{\partial t}(0,t)&= \frac{\partial \sigma_{rr}}{\partial t}(0,t)
\end{align}
\end{subequations}

By using $\frac{\partial v_r}{\partial r}(0,t)\sim\frac{v_r(\frac{1}{2}\Delta r)}{\Delta r/2}$, the following discretized equations are finally obtained for the central point:
\begin{subequations}
\begin{align}
\frac{\sigma_{rr}(0,m+1)-\sigma_{rr}(0,m)}{\Delta t}=& \left[3\lambda(0)+2\mu(0)\right]\frac{1}{\Delta r/2}v_r(\frac{1}{2},m+\frac{1}{2}) - \nonumber \\
&\left[ \lambda(0)+\frac{2}{3}\mu(0)\right]\beta(T(0,m+1/2)) \frac{T(0,m+1)-T(0,m)}{\Delta t} \\
\sigma_{\theta\theta}(0,m)=\sigma_{rr}(0,m)&
\end{align} 
\end{subequations}

The temporal step was to the spatial step by the following stability condition:
\begin{equation}
\Delta t=0.99\times\frac{\Delta r}{\sqrt{3}\; c_{gold}}
\end{equation}
with $c_{gold} = 3240\ \mathrm{m/s}$ the highest speed of sound involved in the problem.

\end{appendix}

\bibliography{PRB_GNS_PA}

\end{document}